%% file: main.tex
\documentclass[10pt,twocolumn,letterpaper]{article}

\usepackage{cvis}              
\usepackage{algorithm}
\usepackage{algorithmic}
\usepackage{amsmath,amssymb}
\usepackage{graphicx}
\input{preamble}
\definecolor{cvisblue}{rgb}{0.21,0.49,0.74}
\usepackage[pagebackref,breaklinks,colorlinks,allcolors=cvisblue]{hyperref}
\def\confName{CVIS}
\def\confYear{2026}

\title{Lightweight Range–Angle Imaging Based Algorithm for Quasi-Static Human Detection on Low-Cost FMCW Radar}

\author{Huy Trinh\\
University of Waterloo\\
{\tt\small h3trinh@uwaterloo.ca}
\and
George Shaker\\
University of Waterloo\\
{\tt\small gshaker@uwaterloo.ca}
}

\begin{document}
\maketitle
\input{sec/0_abstract}    
\input{sec/1_intro}
\input{sec/2_methodologies}
\input{sec/3_results}
\input{sec/4_conclusion}
{
    \small
    \bibliographystyle{ieeenat_fullname}
    \bibliography{main}
}

\end{document}

%% file: sec/0_abstract.tex
\begin{abstract}
Quasi-static human activities such as lying, standing or sitting produce very low Doppler shifts and highly spread radar signatures, making them difficult to detect with conventional constant–false–alarm rate (CFAR) detectors tuned for point targets. Moreover, privacy concerns and low lighting conditions limit the use of cameras in long–term care (LTC) facilities. This paper proposes a lightweight, non-visual image–based method for robust quasi-static human presence detection using a low–cost 60\,GHz FMCW radar.
On a dataset covering five semi-static activities, the proposed method improves average detection accuracy from 68.3\% for Cell-Averaging CFAR (CA-CFAR) and 78.8\% for Order-Statistics CFAR (OS-CFAR) to 93.24\% for Subject 1, from 51.3\%, 68.3\% to 92.3\% for Subject 2, and 57.72\%, 69.94\% to 94.82\% for Subject 3, respectively. Finally, we benchmarked all three detectors across all activities on a Raspberry Pi 4B using a shared Range-Angle (RA) preprocessing pipeline. The proposed algorithm obtains an average 8.2\,ms per frame, resulting in over 120 frames per second (FPS) and a 74$\times$ speed-up over OS–CFAR. These results demonstrate that simple image–based processing can provide robust and deployable quasi-static human sensing in cluttered indoor environments.
\end{abstract}

%% file: sec/1_intro.tex
\section{Introduction}

Semi-static human activities such as lying on a sofa, sitting, or standing are common in long--term--care (LTC) facilities, yet they remain difficult to monitor with conventional radar processing~\cite{10118759,book}. Recent remote-sensing work has focused either on range--Doppler or range--angle feature maps and deep neural networks trained on high-resolution radar data. Christian et al.~\cite{10284529} evaluate several image transformations of range--Doppler maps and study their impact on lightweight convolutional models for person localization, while Stephan et al.~\cite{Stephan2021} propose a deep--learning architecture to detect people in range--Doppler images. These methods, however, do not explicitly target quasi-static occupants with very low radial velocity.
On the other hand, several studies have explored treating radar feature maps as images and applying image processing and morphological operations on range--angle maps to estimate angle--of--arrival and improve visual representation~\cite{electronics10232905,electronics10192397}, and other authors study YOLO--based detectors on range--angle heatmaps to detect multiple targets under various scenarios~\cite{9465137, 10118759}. However, these approaches either rely on heavy deep networks or assume high-end MIMO arrays~\cite{10289261}, which limits their suitability for low-power embedded deployment in LTC rooms.
To the best of our knowledge, no prior work has proposed a lightweight image–based method specifically tailored to quasi–static human presence detection using low-cost FMCW radar with only a few receive channels. In this paper we bridge radar signal processing and image processing by: (i) treating the Capon–based range–azimuth (RA) map as a 2-D image; (ii) systematically evaluating 2-D CA-CFAR and OS-CFAR on these RA images under our real-world high-clutter condition dataset; and (iii) introducing a simple percentile–gated detector on range–angle to improves semi--static detection accuracy and lightweight for edge deployment.

%% file: sec/2_methodologies.tex
\section{Methodologies}
\label{sec:methodologies}
\begin{figure*}[!t]
  \centering    \includegraphics[scale=0.35, trim={0 290 0 0}, clip]{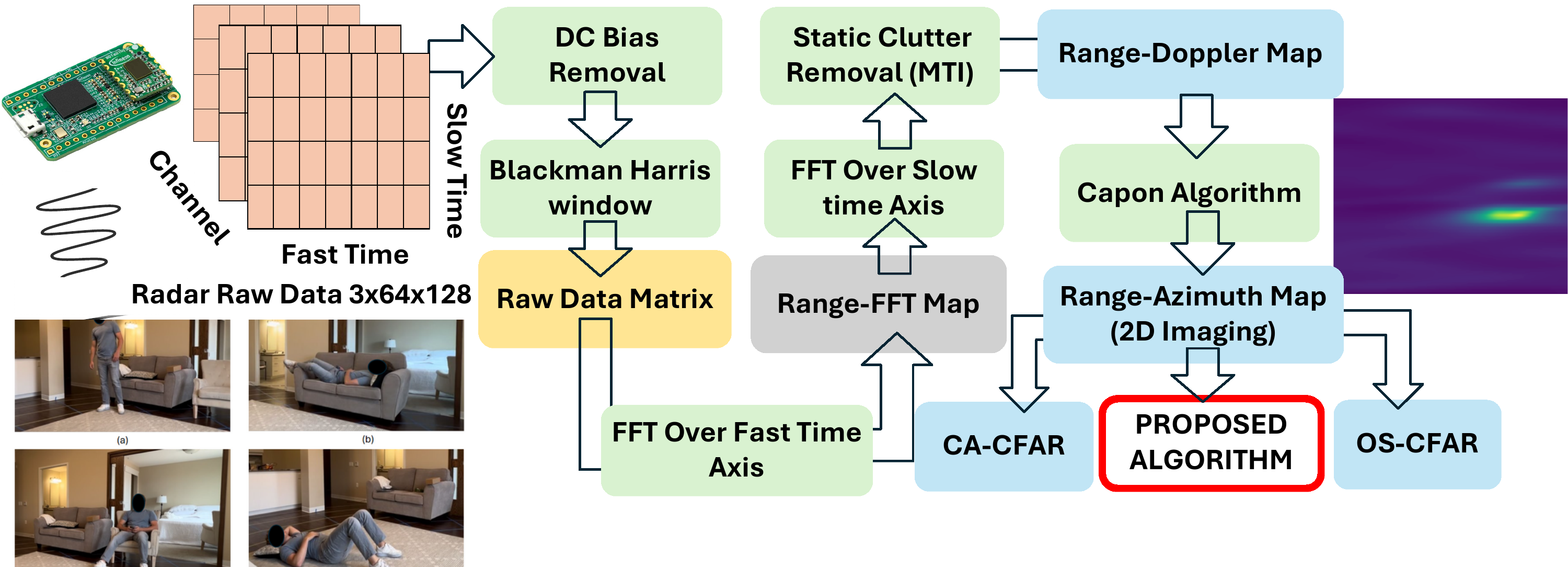}
  \caption{End-to-end radar's imaging processing pipeline. Raw complex IQ frames from the 60\,GHz FMCW radar are converted to range–Doppler maps via range and Doppler FFTs with DC removal, windowing, and MTI clutter suppression. Capon beamforming over two receive channels produces a normalized range–azimuth (RA) intensity map, which is then processed by three detectors: 2-D CA–CFAR, 2-D OS–CFAR, and the proposed percentile-gated lump detector.}
  \label{fig:flow_chart}
\end{figure*}
\subsection{Data Collecting and Preprocessing}
\begin{figure}[htb]
\centerline{\includegraphics[height=0.33\textheight, width=\linewidth]{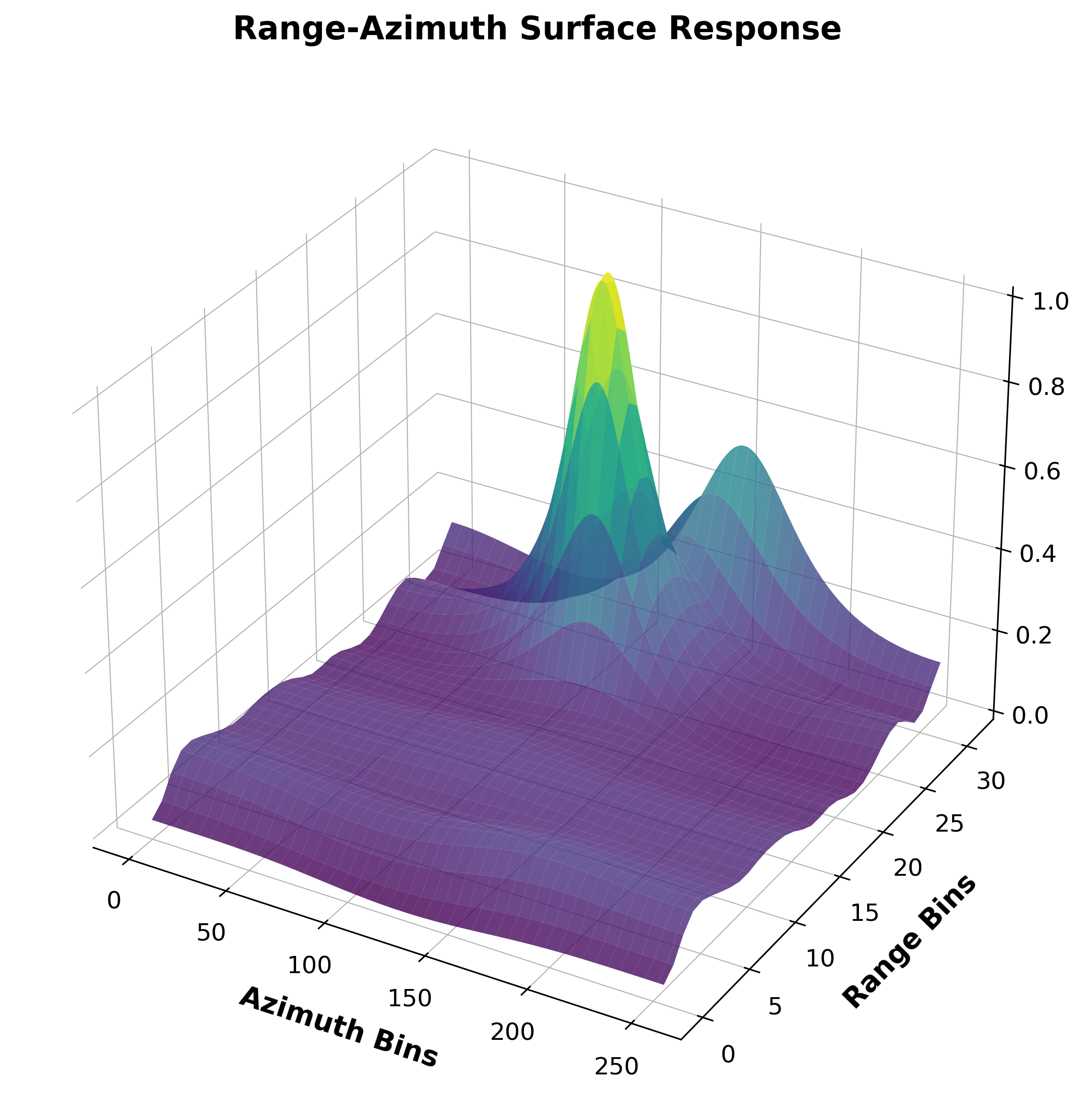}}
\caption{Capon-based range–azimuth surface response (1 frame) for a subject lying on the sofa. The subject produces a compact high-intensity peak in range–azimuth space on top of a broad background clutter ridge from walls and furniture.}
\label{fig:ra_surface}
\end{figure}

For this study, we collected a dedicated dataset of semi-static activities in a LTC using an off--the--shelf 60\,GHz mmWave FMCW radar (Infineon XENSIV\textsuperscript{\texttrademark} BGT60TR13C) with one transmitter and three receivers~\cite{infineon2024bgt60tr13c}. The radar was wall-mounted at approximately $2.5$\,m height with partially overlapping fields of view and a slight downward tilt to cover both sofa and floor areas. The room contains typical living--room furniture and occlusions (sofa, coffee table, TV, windows, and surrounding walls), providing realistic multipath and clutter. For each activity (standing, sitting on the sofa, sitting on floor, lying on the sofa, lying on the floor), three subjects were recorded for two minutes each at multiple aspect angles and ranges at 10\,Hz rate. 
Each radar frame provides a complex--valued data cube of size
$3 \times N_{\text{chirp}} \times N_{\text{sample}} = 3 \times 128 \times 64$.
Our preprocessing pipeline follows Infineon baseline \cite{infineon_an141319} and summarized in Fig.~\ref{fig:flow_chart}. 
To suppress static clutter we apply an exponential moving--average high--pass filter in the RD domain, similar to a moving--target--indicator (MTI) stage~\cite{8703820,trinh2026dopplerdomainrespiratoryamplificationsemistatic}. 
Finally, range-angle feature maps are formed using a Capon beamformer across the two spatially separated receive channels (RX0 and RX2). For each range bin, we estimate a $2\times2$ spatial covariance matrix from Doppler bins and compute the Capon spatial spectrum over azimuth~\cite{10784889,10880536,Abedi2020OnTU}. Stacking the resulting power values over range and azimuth yields the RA map $X \in \mathbb{R}^{H \times W}$ (with $H=64$ ranges and $W=256$ azimuth bins). Each frame is normalized by its maximum value so that $X$ lies in $[0,1]$ as visualized in Fig.~\ref{fig:ra_surface}. These RA images are common inputs to all detection methods used in the remainder of this work.
\subsection{Baseline 2--D CA--CFAR and OS--CFAR}
\begin{figure*}[!t]
  \centering    \includegraphics[scale=0.3, height=0.12\textheight, width=0.8\textwidth]{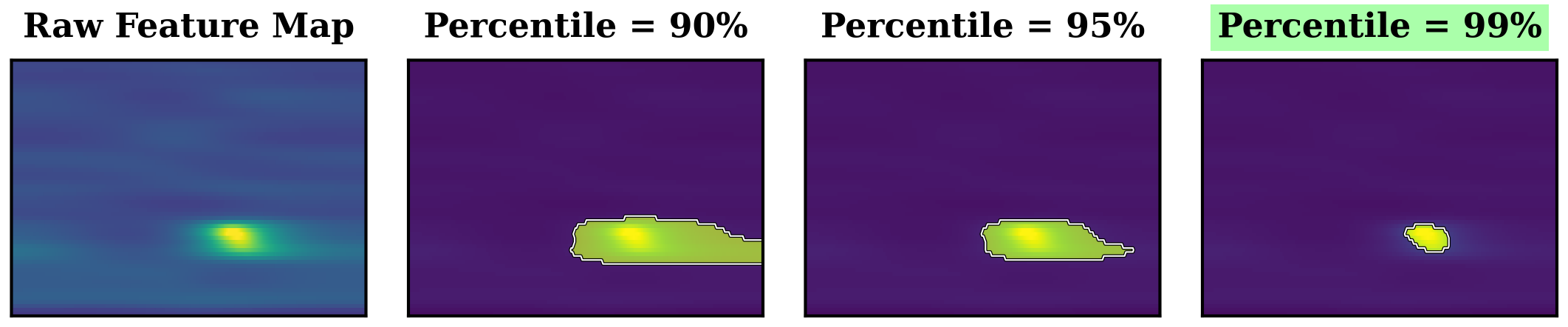}
  \caption{Percentile ablation on a representative range--azimuth frame (\emph{lying on sofa}). From left to right: normalized RA image and detection masks obtained by keeping only the top 90\%, 95\%, and 99\% of pixel intensities.} 
  \label{fig:percentile_ablation}
\end{figure*}
Given a range--azimuth image $X$, conventional 2--D CFAR operates on each cell--under--test (CUT) by comparing its power $|X(\mathrm{CUT})|^2$ to an adaptive threshold derived from neighbouring ``reference'' cells~\cite{10289281,Miller2009FundamentalsOR}. The detection decision can be written as
\begin{equation}
I_{\mathrm{det}}(\mathrm{CUT}) =
\begin{cases}
1, & |X(\mathrm{CUT})|^2 > \mu \hat{\sigma}^2,\\[1ex]
0, & \text{otherwise},
\end{cases}
\label{eq:cfar_basic}
\end{equation}
where $\hat{\sigma}^2$ is an estimate of the local noise--plus--clutter power and $\mu$ is a scale factor chosen to achieve a desired probability of false alarm $P_{\mathrm{fa}}$. In \textbf{cell averaging CFAR (CA--CFAR)}, $\hat{\sigma}^2$ is obtained by averaging the power in a sliding $K \times K$ window of reference cells surrounding the CUT. excluding a guard region directly around the CUT. For a given $P_{\mathrm{fa}}$ and $N$ reference cells, the scale factor is~\cite{8703820}
\begin{equation}
\mu = N \big(P_{\mathrm{fa}}^{-1/N} - 1\big),
\label{eq:ca_cfar_mu}
\end{equation}
and the threshold $\mu \hat{\sigma}^2$ is updated at each pixel as the window moves across the RA image. \textbf{Order statistics CFAR (OS--CFAR)} \cite{4102829} replaces the mean of the reference window by an order statistic that is more robust to outliers: the reference powers are sorted and the $k$-th largest value is used as $\hat{\sigma}^2$~\cite{8703820}.
\subsection{Proposed Range--Angle Subject Detector} 
Instead of applying CFAR directly on the RA image, we treat $X$ as a 2-D grayscale image and search for a single connected lump that corresponds to the human reflector. 
Our detector focuses on the top $p$-percentile pixel intensities and enforces spatial coherence. 

\begin{algorithm}[thb]
\caption{Range–Angle Lump Detector}
\label{alg:ra_lump_simple}
\begin{algorithmic}[1]
\STATE $\tau \gets \textsc{Percentile}(X, p)$ \COMMENT{gate top-$p$ intensity}
\STATE $M \gets 0_{H\times W}$
\FORALL{pixels $(i,j)$}
   \IF{$X[i,j] \ge \tau$}
      \STATE $M[i,j] \gets 1$ \COMMENT{keep bright responses}
   \ENDIF
\ENDFOR
\STATE $M \leftarrow \mathrm{Close}(M)$ \COMMENT{connect nearby pixels (small structuring element)}
\STATE $M \leftarrow \mathrm{RemoveSmall}(M, A_{\min})$ \COMMENT{discard tiny blobs}
\STATE $\mathcal{C} \leftarrow \mathrm{ConnectedComponents}(M)$
\IF{$|\mathcal{C}| = 0$} \RETURN $M\!=\!0_{H\times W},\, B\!=\!\emptyset$ \ENDIF
\STATE $c^\star \leftarrow \arg\max_{c\in\mathcal{C}} \mathrm{Area}(c)$ \COMMENT{dominant lump}
\STATE $M \gets 0_{H\times W}$;
       \FORALL{pixels $(i,j)$ in $c^*$} \STATE $M[i,j] \gets 1$ \ENDFOR
\STATE $B \gets \textsc{BBox}(c^*)$
\STATE \RETURN $M, B$
\end{algorithmic}
\end{algorithm}
Algorithm~\ref{alg:ra_lump_simple} summarizes our proposed procedure. Given a normalized RA image $X$, we first compute a global percentile threshold $\tau$ and form a binary mask $M$ that keeps only pixels above $\tau$. This step removes most of the floor and wall clutter while preserving the main lobe of the human response. Next, we apply a small $2\times2$ structuring element for binary closing to connect nearby bright pixels along range and azimuth, and we remove any blobs whose area is below $A_{\min}$. A two–pass connected–components algorithm labels the remaining blobs, and we keep only the largest component as the final detection lump. 
All operations in Algorithm~\ref{alg:ra_lump_simple} are linear in the number of pixels and involve only simple comparisons and logic, resulting in $\mathcal{O}(HW)$ complexity per frame. In contrast, 2-D CA-CFAR and OS-CFAR require sliding windows over each pixel with relatively large kernels 
, resulting in $\mathcal{O}(HWK^{2})$ complexity and substantially higher runtime. 
Figure~\ref{fig:percentile_ablation} illustrates that different percentile intensity $p$ affects the detection mask on a representative RA frame for the lying on sofa activity of subject 1. 
At the 90th and 95th percentiles, the mask still includes a wide horizontal band along the dominant clutter ridge, causing the detected lump to spread across the entire sofa region. In contrast, the 99th percentile isolates a smaller, well-localized lump that aligns with the annotated human activity detection.

%% file: sec/3_results.tex
\section{Results and Discussion}
\label{sec:results}
\begin{figure}[t]
  \centering    \includegraphics[scale=0.24]{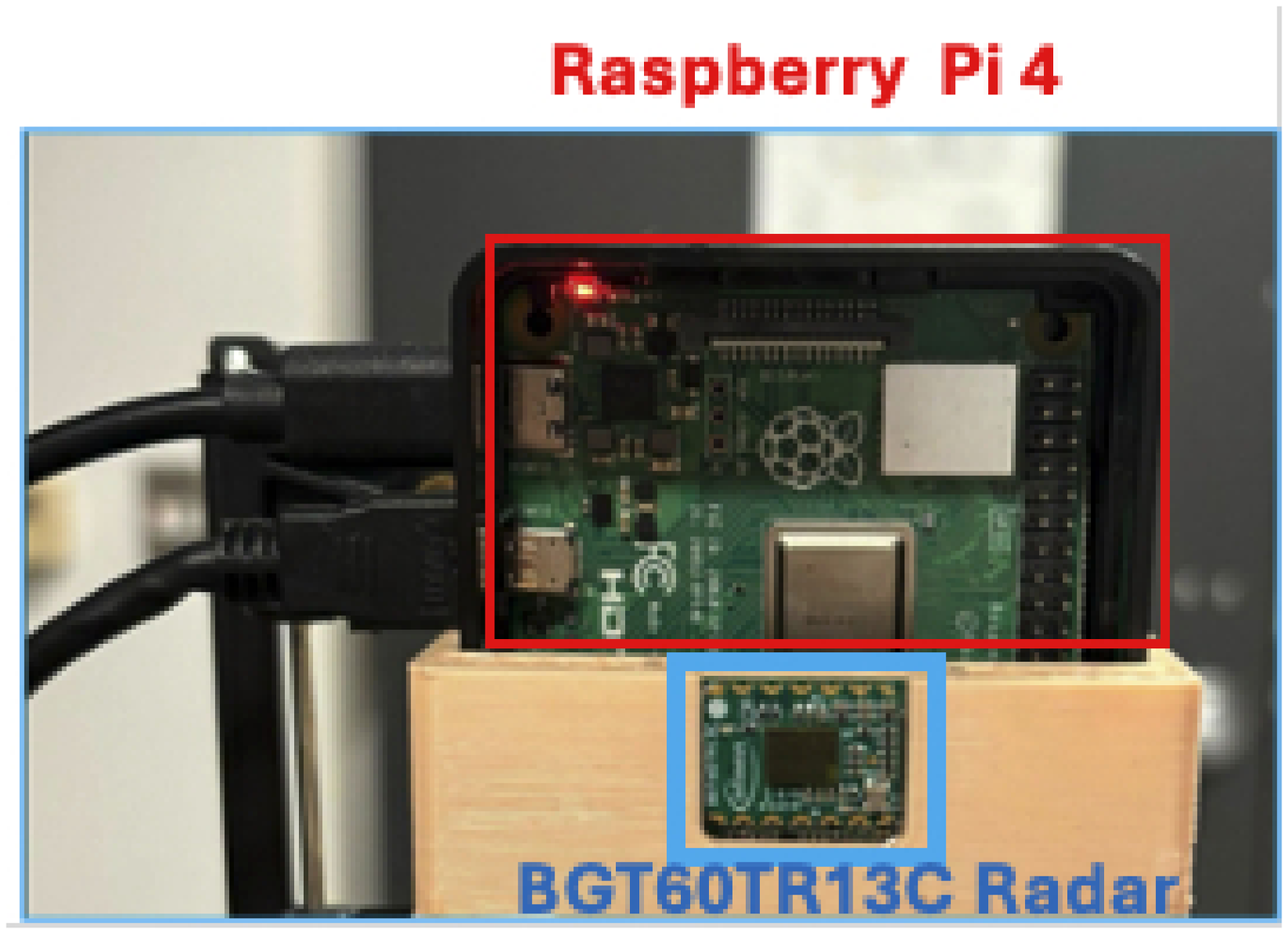}
  \caption{Raspberry Pi 4B Board with BGT60TR13C radar.}
 \label{fig:pi_and_radar}
\end{figure}

\begin{table*}[t]
\centering
\caption{Detection accuracy (\%) per subject. Best values for each metric per subject are shown in \textbf{bold}.}
\label{tab:accuracy}
\setlength{\tabcolsep}{11pt} 
\renewcommand{\arraystretch}{1.15}
\begin{tabular}{l ccc | ccc | ccc}
\hline
& \multicolumn{3}{c}{\textbf{Subject 1}} & \multicolumn{3}{c}{\textbf{Subject 2}} & \multicolumn{3}{c}{\textbf{Subject 3}} \\
\textbf{Activity} & CA & OS & \textbf{Ours} & CA & OS & \textbf{Ours} & CA & OS & \textbf{Ours} \\
\hline
Standing     & 79.0 & 89.9 & \textbf{93.5} & 45.0 & 65.7 & \textbf{91.7} & 80.0 & 91.2 & \textbf{97.3} \\
Sit on Sofa  & 75.7 & 88.2 & \textbf{97.0} & 31.4 & 51.6 & \textbf{86.9} & 57.0 & 72.2 & \textbf{92.7} \\
Sit on Floor & 83.3 & 91.5 & \textbf{97.2} & 54.6 & 68.4 & \textbf{94.1} & 74.7 & 82.9 & \textbf{97.4} \\
Lay on Floor & 70.6 & 80.5 & \textbf{91.9} & 63.9 & 77.0 & \textbf{94.7} & 78.4 & 86.4 & \textbf{93.7} \\
Lay on Sofa  & 33.0 & 43.9 & \textbf{86.6} & 61.8 & 79.0 & \textbf{94.1} & 48.5 & 67.0 & \textbf{93.0} \\
\hline
\textit{Average} & 68.3 & 78.8 & \textbf{93.24} & 51.3 & 68.3 & \textbf{92.3} & 57.72 & 69.94 & \textbf{94.82} \\
\hline
\end{tabular}
\end{table*}
\begin{figure}[htb]
  \centering    \includegraphics[scale=0.25]{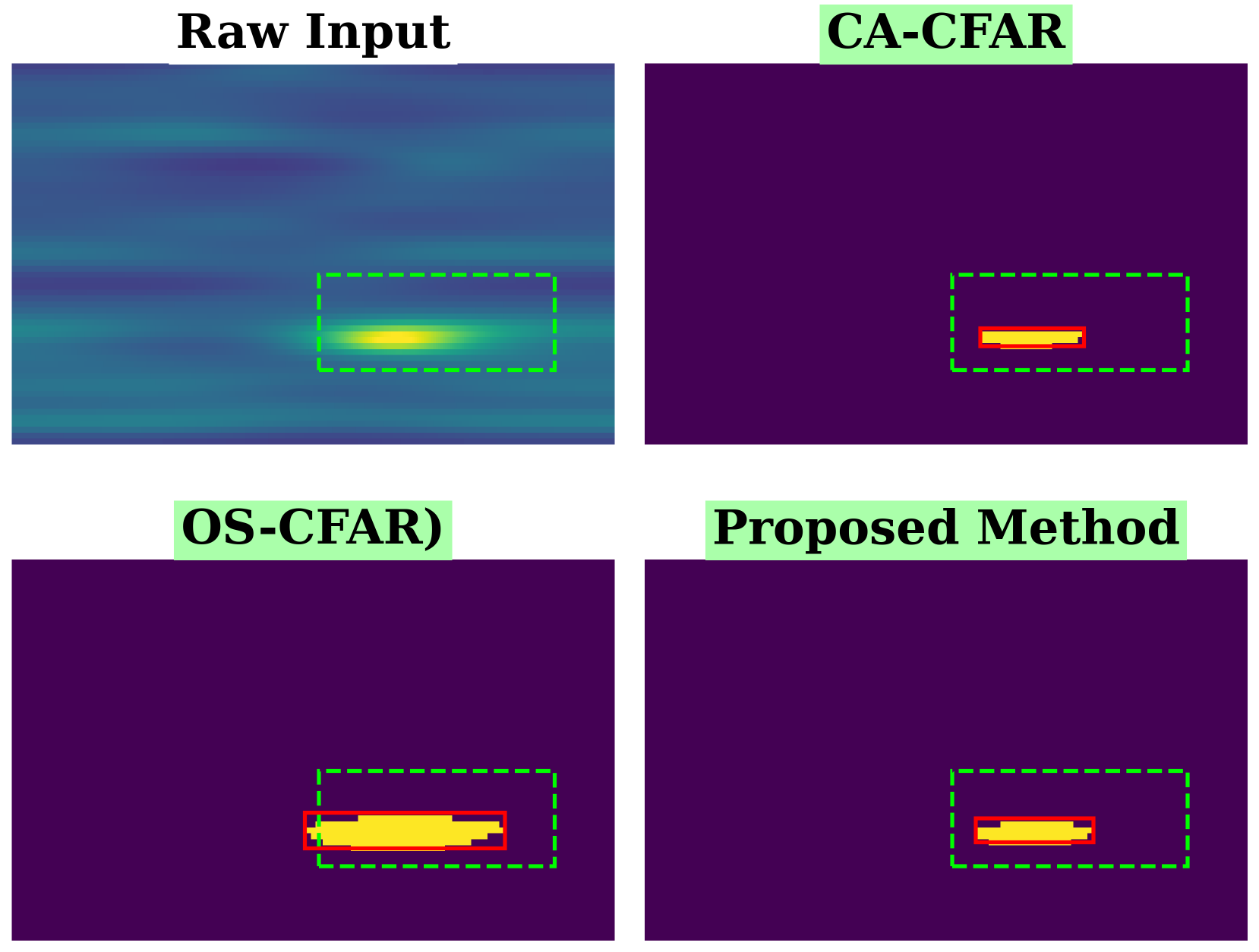}
  \caption{Qualitative comparison of all three detectors on frame 21 for subject 1 lying on the floor, which all of them produce detections within the ground-truth bounding box.}
 \label{fig:Case3_AllPass_Frame_21}
\end{figure}
\begin{figure}[htb]
  \centering    \includegraphics[scale=0.25]{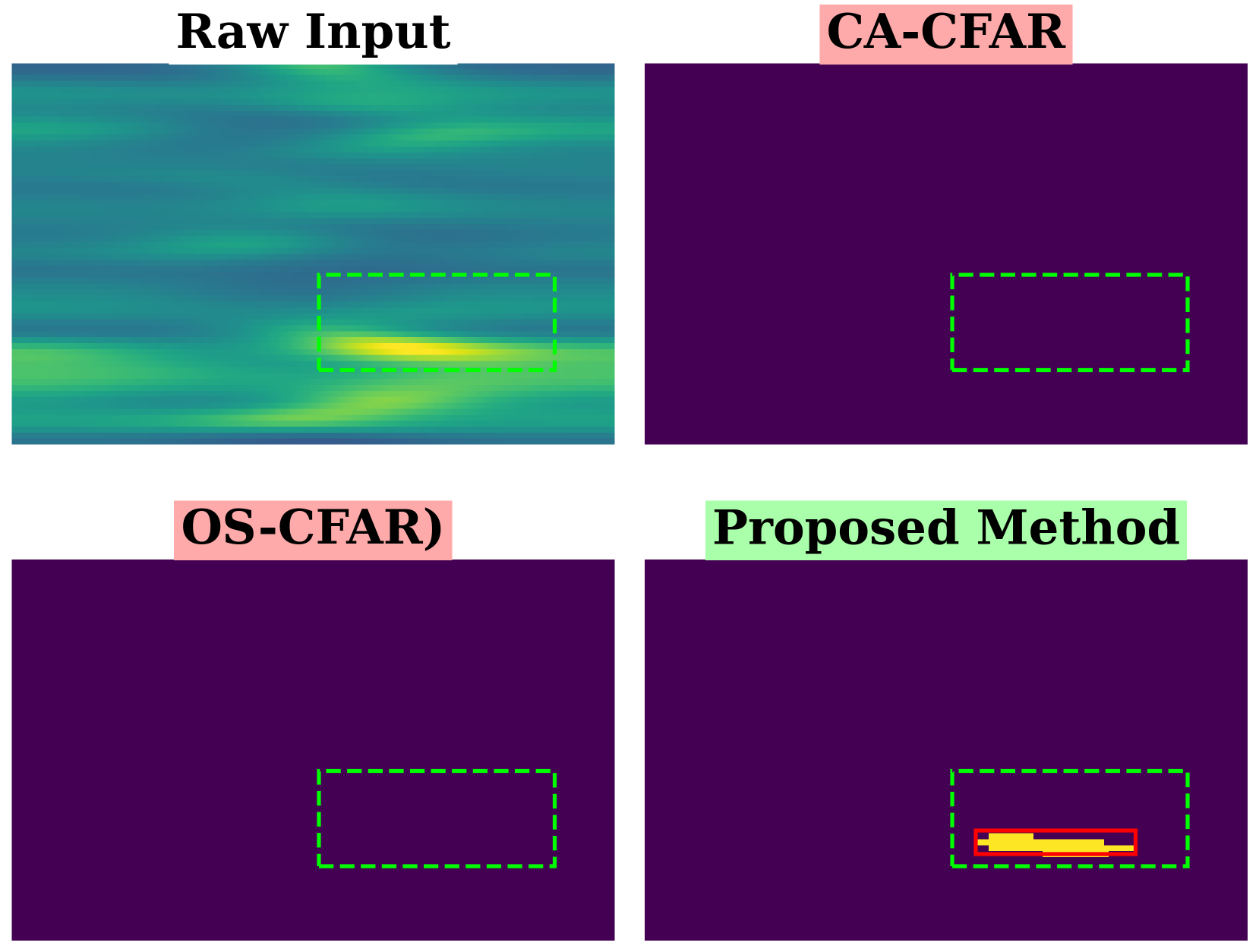}
  \caption{Qualitative comparison of all three detectors on frame 106 for subject 1 lying on the floor. only our proposed method can detect within the ground-truth bounding box.}
  \label{fig:Case2_PropOnly_Frame_106}
\end{figure}
Figures~\ref{fig:Case3_AllPass_Frame_21} and~\ref{fig:Case2_PropOnly_Frame_106} compare the detection results on representative frames for Subject 1 lay on sofa. In these visualizations, the subject's actual position is annotated by a green dashed bounding box. We further encode the decision outcome in the method headers:  we count a frame as a hit if one detected blob overlaps the annotated ground–truth bounding box. A green highlight indicates a successful detection, while a red highlight signifies a miss. While all methods successfully localize the subject within the ground-truth region during the high signal-to-noise ratio (SNR) scenario (Frame 21, Fig.~\ref{fig:Case3_AllPass_Frame_21}), the proposed algorithm demonstrates superior stability under severe noise. As shown in Frame 106 (Fig.~\ref{fig:Case2_PropOnly_Frame_106}), strong background clutter causes both CA–CFAR and OS–CFAR to fail, whereas our proposed method succeeds in detecting the subject within the ground-truth bounds. Table~\ref{tab:accuracy} reports per–activity detection accuracy for three subjects. On average, the proposed method improves detection accuracy to 93.24\% (Subject 1), 92.3\% (Subject 2), and 94.82\% (Subject 3), significantly outperforming CA-CFAR (68.3\%/51.3\%/57.72\%) and OS-CFAR (78.8\%/68.3\%/69.94\%). 
and 43.9\% (OS) to 86.6\% for Subject~1, and 48.5\% (CA),67.0\%(OS) to 93\% for Subject~3. For Subject 2, the increase is from 31.4\% and 51.6\% to 86.9\% for Subject~2 in Sit on Sofa. Notably, sofa-based activities remain the most difficult, representing the lowest detection performance for Subject~1 (lying) and Subject~2 (sitting), respectively. Finally, to assess real-world feasibility and performance, we benchmark all three detectors across all activities using a streaming implementation on a Raspberry Pi 4B \cite{raspberrypi2024pi4b} as Fig.~\ref{fig:pi_and_radar}, and the results are shown in Table~\ref{tab:edge_runtime_compact}. Range–azimuth feature maps are precomputed using the shared pipeline and then fed to each detector. Compared to CA-CFAR (91.2 ms) and OS-CFAR (601.1 ms), our proposed method runs on average of just 8.2 ms per frame, achieving a throughput of over 122 FPS. This represents a $74\times$ speed-up over OS-CFAR and ensures the system easily meets the 10 Hz radar frame-rate constraint without deadline misses. These results confirm that the proposed image-based approach delivers both robust detection accuracy and real-time computational efficiency, suitable for the low-latency requirements of edge deployment.
\begin{table}[thb]
\centering
\caption{Average runtime analysis on Raspberry Pi 4 for 3 methods. Best values for each metric are shown in \textbf{bold}.}
\label{tab:edge_runtime_compact}
\setlength{\tabcolsep}{10pt} 
\renewcommand{\arraystretch}{1.1}
\resizebox{\linewidth}{!}{%
\begin{tabular}{l c c c c}
\toprule
Method & Latency & FPS & RAM & Speedup \\
 & (ms) & & (MB) & (vs. OS) \\
\midrule
CA-CFAR & 91.2 & 11.0 & 299.4 & $6.6\times$ \\
OS-CFAR & 601.1 & 1.7 & 299.5 & $1.0\times$ \\
\textbf{Ours} & \textbf{8.2} & \textbf{122.4} & 299.5 & \textbf{73.6$\times$} \\
\bottomrule
\end{tabular}%
}
\end{table}

%% file: sec/4_conclusion.tex
\section{Conclusion and Future Work}
This paper presented a non-visual sensing solution for quasi-static human presence in cluttered indoor environments using a single low-cost 60 GHz FMCW radar. By treating Capon-based range-azimuth maps as 2-D images and applying a lightweight percentile-gated detector, we achieved robust detection where conventional CFAR methods struggle, particularly in high-clutter scenarios. The method relies on basic image-processing primitives, ensuring it is computationally efficient for real-time execution on embedded platforms. Future work will investigate deeper optimization of the processing pipeline to further enhance detection reliability and stability. A key focus will be on the scalability of this edge-based solution. Given the low computational footprint and hardware cost, we aim to validate the system's deployment in large-scale facilities—scaling to scenarios in hospitals and long-term care centers, where privacy-preserving, decentralized, and cost-effective monitoring are critical.